\documentclass[a4paper,12pt]{article}
\pdfoutput=1
\usepackage[margin=1in]{geometry}
\usepackage{graphicx,xcolor}
\usepackage{mathtools,braket,blkarray}
\allowdisplaybreaks
\usepackage{amssymb,amsfonts,bbm,relsize}
\usepackage[width=0.9\textwidth,footnotesize]{caption}
\usepackage[font=scriptsize]{subcaption}
\usepackage{cite,hyperref,url}
\hypersetup{pageanchor=false}
\usepackage{booktabs,multirow,makecell}
\usepackage[utf8]{inputenc}

\newcommand{\overbar}[1]{\mkern 1.5mu\overline{\mkern-1.5mu#1\mkern-1.5mu}\mkern 1.5mu}

\begin{document}

\begin{titlepage}
\begin{center}
{\bf\Large An $S_4\times SU(5)$ SUSY GUT of flavour in 6d } \\[12mm]
Francisco~J.~de~Anda$^{\dagger}$%
\footnote{E-mail: \texttt{franciscojosedea@gmail.com}},
Stephen~F.~King$^{\star}$%
\footnote{E-mail: \texttt{king@soton.ac.uk}}
\\[-2mm]

\end{center}
\vspace*{0.50cm}
\centerline{$^{\star}$ \it
School of Physics and Astronomy, University of Southampton,}
\centerline{\it
SO17 1BJ Southampton, United Kingdom }
\vspace*{0.2cm}
\centerline{$^{\dagger}$ \it
Tepatitl{\'a}n's Institute for Theoretical Studies, C.P. 47600, Jalisco, M{\'e}xico}
\vspace*{1.20cm}

\begin{abstract}
{\noindent We propose a 6d model with a SUSY $SU(5)$ gauge symmetry. After compactification, it 
explains the origin of the $S_4$ Family Symmetry with CSD3 vacuum alignment, as well as $SU(5)$ breaking with 
doublet-triplet splitting. The model naturally accounts for 
all quark and lepton (including neutrino) masses and mixings, incorporating the highly predictive Littlest Seesaw structure.
It spontaneously breaks CP symmetry, resulting in successful CP violation in the quark and lepton sectors, while
solving the Strong CP problem. It also explains the Baryon Asymmetry of the Universe (BAU) through leptogenesis,
with the leptogenesis phase directly linked to the Dirac and Majorana phases.}
\end{abstract}
\end{titlepage}

\section*{Introduction}
The origin of the three families of quarks and leptons remains a puzzle of the Standard Model (SM), as does
their pattern of masses and mixing parameters.
The SM does not specify the origin of neutrino mass, which, together with lepton mixing and CP violation, introduces a further 9 undetermined parameters. 
The flavour problem, enriched by neutrino oscillation physics, is therefore a major motivation for going beyond the SM.

Clues to a theory of flavour may be found in the neutrino sector.
While the neutrino masses are very small, the lepton mixing angles are rather large, with the approximate tri-bimaximal structure of the
PMNS matrix suggesting some non-Abelian discrete Family Symmetry which admits triplet representations
such as $A_4$ or $S_4$ \cite{Ishimori:2010au}.

Another motivation for going beyond the SM is the quest for unification of matter and gauge forces, as in Grand Unified Theories (GUTs)
\cite{Langacker:1980js}.
To maintain the mass hierarchy in a natural way in GUTs
requires supersymmetry (SUSY), which also facilitates gauge coupling unification \cite{Chung:2003fi}. 
The most ambitious such theories also combine Family Symmetry with GUTs  \cite{King:2017guk}. 
Clearly, if quarks and leptons are unified, then any Family Symmetry introduced in the lepton sector would also
govern the quarks. 

Following the path of 
minimality, one is quickly led to consider 
the minimal simple GUT group, namely $SU(5)$ \cite{Georgi:1974sy}. Since the lepton mixing PMNS matrix arises from the left-handed 
lepton doublets, one is led to unify the $\overline{5}$ representations into a triplet $3$ of the Family Symmetry. 
Moreover, since the top quark has an order unity
Yukawa coupling, one is led to assign the $10$ representations to singlets of the Family Symmetry \cite{Hagedorn:2010th}.

Even such a minimal $SU(5)$ GUT group requires a whole sector to break the gauge symmetry to the SM and to achieve the doublet-triplet Higgs splitting of the light electroweak doublet from the heavy colour triplet \cite{Antusch:2014poa,Bjorkeroth:2015ora}. 
However many of the complications of doublet-triplet splitting are avoided by assuming the existence of extra dimensions \cite{Kawamura:2000ev}.

We are thus led to the notion of Family Symmetry with SUSY GUTs in extra dimensions, for example based on 
$A_4 \times SU(5)$ \cite{Altarelli:2008bg,Burrows:2009pi,Burrows:2010wz}.
\footnote{An additional $U(1)$ Family Symmetry is
sometimes assumed in order to yield hierarchies between different families \cite{Altarelli:2000fu} via the Froggatt-Nielsen mechanism
\cite{Froggatt:1978nt}. In our model we shall also assume an additional global $U(1)$ as a ``shaping symmetry'' to control the non-renormalisable
operator structure.} In such theories, the discrete Family Symmetry could have a dynamical origin as a result
of the compactification of a 6d theory down to 4d
\cite{Altarelli:2006kg,Adulpravitchai:2010na,Adulpravitchai:2009id,Asaka:2001eh}.
The connection to string theory of these and other orbifold
compactifications has also been discussed in \cite{Kobayashi:2006wq}.

In this paper we propose a model based on $S_4 \times SU(5)$ in extra dimensions.
We generate the $S_4$ symmetry from the orbifold, by a generalisation of the mechanism previously used to obtain $A_4$. The
motivation for considering the group $S_4$ is that it is better suited to yielding the so called ``CSD3'' vacuum alignments \cite{King:2013iva} that we desire,
leading to the highly predictive Littlest Seesaw structure \cite{King:2015dvf}, with 3 input parameters predicting 9 observables in the neutrino sector,
which agrees perfectly with current neutrino oscillation measurements \cite{Ballett:2016yod}. 
With the help of the orbifold conditions, we shall show that obtaining this alignment can be much simpler than just using a 4d superpotential.

The aim of this paper, then, is to propose a 6d model with a SUSY $SU(5)$ gauge symmetry that automatically results in an $S_4$ 
Family Symmetry after compactification, where the neutrinos arising from the CSD3 vacuum alignment have the Littlest Seesaw structure. 
Our goal is to construct a fairly
complete and natural model that fits all flavour observables while being highly predictive in the neutrino sector. 

After compactification,
and GUT and flavour symmetry breaking, the low energy theory is just the 
Minimal Supersymmetric Standard Model (MSSM), supplemented by two extra Right Handed Neutrinos (RHNs),
however with a constrained set of input parameters.
In particular: there is only a single input phase parameter in each of the lepton and quark sectors;
the Littlest Seesaw structure arises in the neutrino sector; the Yukawa matrices in the down/charged lepton sectors
have upper/lower triangular form,  on the one hand resulting in a solution to the Strong CP problem and, on the other hand,
very small charged lepton mixing contributions to the PMNS matrix. Charged fermion masses are naturally hierarchical,
due to a hierarchy of flavon VEVs,
while the neutrino masses have normal ordering with the lightest neutrino mass being zero.

The main successes of the resulting SUSY $SU(5)$ model in 6d are summarised as follows:
\begin{itemize}
\item{It explains, after compactification, the origin of the $S_4$ Family Symmetry with CSD3 vacuum alignment, as well as $SU(5)$ breaking with 
doublet-triplet splitting.}
\item{It naturally accounts for 
all quark and lepton (including neutrino) masses and mixings, incorporating the highly predictive Littlest Seesaw structure.}
\item{It spontaneously breaks CP symmetry, resulting in successful CP violation in the quark and lepton sectors, while
solving the Strong CP problem.}
\item{It also explains the Baryon Asymmetry of the Universe (BAU) through leptogenesis,
with the leptogenesis phase directly linked to the Dirac and Majorana phases.}
\end{itemize}

The layout of the paper is as follows. In sec. \ref{sec:model} we present the field content of the model. In sec. \ref{sec:orbifold} we show the details of the orbifold, focusing on how $S_4$ is generated through this orbifolding. In sec. \ref{sec:flaval} we show how the flavon VEVs are fixed in to the CSD3 alignment. In \ref{sec:smfer}, the fermion mass matrices and a numerical fit are presented. In sec. \ref{sec:strcp} we show how the model solves naturally the Strong CP problem. In sec. \ref{sec:leptg} we show how the BAU can be naturally obtained through leptogenesis. In sec. \ref{sec:prodec} we show how proton decay is controlled.  In the appendix \ref{app:6dpot} we show the original 6d superpotential before compactification. Finally, in appendix \ref{app:discrete} we show how the same model can be obtained with a discrete shaping symmetry instead of the $U(1)$.

\section{Overview of the model}
\label{sec:model}

The model is based on a six dimensional spacetime with $\mathcal{N}=1$ supersymmetry (SUSY), an $SU(5)$ gauge symmetry and a global $U(1)$ symmetry
that we refer to as a ``shaping'' symmetry, since it governs the allowed operators.
The extra dimensions are compactified in a torus orbifold $T^2/(\mathbb{Z}_2^{SM}\times \mathbb{Z}_2)$. This compactification breaks the extended SUSY and the GUT groups. This orbifolding is done in a standard way and it is summarized in the appendix \ref{sec:orbifold}. 

The way the compactification is done leaves a remnant $S_4$ symmetry which we identify as the flavour group. The fields can be chosen so that, after the compactification, they transform under irreducible representations of this $S_4$. The 4 fixed branes are related by $S_4$ transformations and so are the fields located in them. Since the 4 branes are interchanged by $S_4$, we will simply refer to them as the brane. This is shown in appendix \ref{sec:s4orbi}. 

Besides the gauge superfields, the model contains the chiral superfields that are listed in table \ref{tab:funfields}. There we list the representation of each field under the GUT group $SU(5)$, the flavour group $S_4$ and their charges under the shaping symmetry $U(1)$. If the field propagates in the bulk, it should be an eigenstate of the boundary condition matrix $P'_{SM}$ and its parity $\pm 1 $ is listed in the last column. If the field is located in the brane, it is stated as Brane in the last column.

\begin{table}
\centering
\footnotesize
\captionsetup{width=0.9\textwidth}
\begin{tabular}{| c | c@{\hskip 5pt}c |  c c|}
\hline
\multirow{2}{*}{\rule{0pt}{4ex}Field}	& \multicolumn{4}{c |}{Representation} \\
\cline{2-5}
\rule{0pt}{3ex}			& $S_4$ & $SU(5)$ &  $U(1)$ & $P_{SM}'$ \\ [0.75ex]
\hline \hline
\rule{0pt}{3ex}%
$F$ 			& 3' & $\bar{5}$  &  $-c$ & Brane \\
$T_1^\pm$ 			& 1& $10$ & $a-4d$ & $\pm 1$\\
$T_2^\pm$ 			& 1& $10$ & $a-2d$ & $\pm 1$\\
$T_3^\pm$ 			& 1& $10$  & $a$ & $\pm 1$\\
$N_s^c$ 			& 1& $1$  & $-d$ & $+1$\\
$N_a^c$ 			& 1& $1$ & $-4d$ & $+1$\\
$H_5$ 			& 1& $5$  & $-2a$ & $+1$\\
$H_{\bar{5}}$ 			& 1& $\bar{5}$ & $-2b$ & $+1$\\
\rule{0pt}{3ex}%
$\xi$ 			& 1& 1 & $2d$ & $+1$\\
$\rho$ 			& $2$& $1$ & $-a+2b+c+d$ & $+1$\\
$\phi_s$ 			& 3'& 1  & $2a+c+d$ & Brane \\
$\phi_a$ 			& 3'& 1  & $2a+c+2d$ & $-1$\\
$\phi_\tau$ 			& 3'& 1 & $-a+2b+c$ &Brane \\
$\phi_\mu$ 		& 3'& 1 & $-a+2b+c+2d$ &Brane \\
$\phi_e$ 			& 3'& 1 & $-a+2b+c+4d$ & $+1$\\
\rule{0pt}{3ex}%
$A_1$ 			& 1& 1 &  $2a-4b-2c$ & $+1$\\
$A_{3'}$ 			& 3'& 1 &  $-a - 2b - 2 c - 2 d$ &Brane \\
$A_{2}$ 			& 2& 1 &  $2 a - 4 b - 2 c - 8 d$ & $+1$\\
$A_{1}'$ 			& 1'& 1 &  $2 a - 4 b - 2 c - 4 d$ &Brane \\
\hline
\end{tabular}
\caption{Complete list of chiral superfields in the model.  A setup that gives exactly the desired Yukawa terms would be with $\{a,b,c,d\}=\{7,13,1,2\}$.}
\label{tab:funfields}
\end{table}

The SM fermions lie inside the $F, T_i$ as in an usual $SU(5)$ theory. The flavour triplet $F$ contains $d^c$ and $L$. There are two copies of each $T_{i}$ in the bulk, each with different parity under the $\mathbb{Z}_2^{SM}$ boundary condition. The $T^+$ contains $u^c$ and $e^c$, while the $T^-$ contains $Q$. This allows different masses for charged leptons and down quarks. We have only two right handed neutrinos (RHN) $N^c_{a,s}$, as this is the minimum case. The MSSM doublets $h_{u,d}$ lie in $H_{5,\bar{5}}$ respectively. They are located in the bulk with a positive parity so that only the doublets are light after compactification, so that the doublet triplet splitting is natural.

The superfields $\xi,\rho,\phi$ are flavons that give structure to the SM fermion mass matrices. Some of them lie in the bulk and some in the brane, depending on their alignment. The fields $A_a$ are alignment fields and whose F-term equations fix the alignment of the flavons. 

We assume that compactification happens at the GUT scale so that gauge coupling unification happens naturally and every KK mode and every extra field is located at the GUT scale. Furthermore the compactification gives the flavour symmetry and helps to align the flavons that break it. At low energies, we have exactly the MSSM. 

As we shall see the model is rather complete and predictive. Models that aim to be complete usually end up with a very large number of fields to achieve it \cite{Bjorkeroth:2015ora, Bjorkeroth:2015uou, Bjorkeroth:2017ybg,deAnda:2017yeb}.
 In this model, the extra dimensions play a big role in achieving symmetry breaking, so that the full field content, listed in table \ref{tab:funfields}, is much smaller than any previous theory.  

\section{Orbifolding}
\label{sec:orbifold}

We are considering a $\mathcal{N}=1$ supersymmetric Yang-Mills
theory in 6 dimensions, the Lagrangian reads,

\begin{equation}
\mathcal{L}^{YM}_{6d}=
\mathrm{Tr}\left(-\frac{1}{2}V_{MN}V^{MN}+i\overline{\Lambda}\Gamma^MD_{M}\Lambda\right),
\end{equation}
where $V_{M}= t^aV^{a}_{M}$ and $\Lambda = t^a\Lambda^a$, here
$t^a$ are the generators of SU(5). $D_M\Lambda=
\partial_m\Lambda-ig[V_M,\Lambda]$ and $V_{MN}= [D_M,D_N]/(ig)$.
The $\Gamma$ matrices are given by:
\begin{align}
\Gamma^\mu =
 \begin{pmatrix}
   \gamma^\mu & 0 \\
   0 & \gamma^\mu
 \end{pmatrix},\qquad
 \Gamma^5 =
 \begin{pmatrix}
   0 & i\gamma^5  \\
  i\gamma^5 & 0
 \end{pmatrix},\qquad
\Gamma^6 =
 \begin{pmatrix}
   0 & \gamma^5 \\
  -\gamma^5 & 0
 \end{pmatrix}
\end{align}
with $\gamma^5 = I$ and $\{ \Gamma_M, \Gamma_N
\}=2\eta_{MN}\mathbf{1}_{(8\times 8)},\eta_{MN}=\mathrm{diag}(1,-1,-1,-1,-1,-1).$ The gaugino
$\Lambda$ is composed of two Weyl fermions of opposite chirality
in 4d,
\begin{eqnarray}
 \Lambda=(\lambda_1, -i\lambda_2), &\gamma_5\lambda_1=-\lambda_1,& \gamma_5\lambda_2=\lambda_2.
\end{eqnarray}
Overall the gaugino has negative 6d chirality
$\Gamma_7\Lambda=-\Lambda$, where
$\Gamma_7=\mathrm{diag}(\gamma_5,-\gamma_5).$

We assume that the spacetime is $\mathcal{M}=\mathbb{R}^4\times \mathbb{T}^2$, where the torus is defined by 
\begin{equation}\begin{split}
(x^5,x^6)&=(x^5+2\pi R_1,x^6),
\\ (x^5,x^6)&=(x^5+2\pi R_2\cos\theta,x^6+2\pi R_2\sin\theta).
\label{eq:tras}
\end{split}\end{equation}
Where $R_{1,2}$ are the radii of the extra dimensions and define the compactification scale
\begin{equation}
M_C\sim \frac{1}{R_{1,2}}.
\end{equation}

The orbifold is chosen to be $\mathbb{T}^2/(\mathbb{Z}_2\times\mathbb{Z}_2^{SM})$.

The first $\mathbb{Z}_2$ orbifolding comes from the parity condition 
\begin{equation}
\mathbb{Z}_2:\ \ \ (x^5,x^6)=(-x^5,-x^6),
\label{eq:par}
\end{equation}
applied the vector supermultiplet:
\begin{eqnarray}
 PV_\mu(x,-x_5,-x_6)P^{-1} & = & +V_\mu(x,x_5,x_6),\\
PV_{5,6}(x,-x_5,-x_6)P^{-1} & = & -V_{5,6}(x,x_5,x_6),
\end{eqnarray}
and
\begin{eqnarray}
 P\lambda_1(x,-x_5,-x_6)P^{-1}&=&+\lambda_1(x,x_5,x_6),\\
P\lambda_2(x,-x_5,-x_6)P^{-1}&=&-\lambda_2(x,x_5,x_6).
\end{eqnarray}

This also happens if we locate chiral 6d supermultiplets in the bulk. They would decompose into two 4d fermions with opposite parities, a complex 4d scalar and a complex 4d pseudoscalar. These are arranged into 2 chiral 4d supermultiplets as usual.

With $P=I$, the effective
$\mathcal{N}=2$ SUSY in 4d is broken down to $\mathcal{N}=1$.

The second orbifolding is done at
\begin{equation}
(x'_5,x'_6)=(x_5+\pi R_1,x_6),
\end{equation}
with the condition
\begin{equation}
\mathbb{Z}^{\mathrm{SM}}_2 :\ \ \ (x'_5,x'_6)=(-x'_5,-x'_6).
\end{equation}
Now the condition would be 
\begin{equation}
 P_{SM}=\begin{pmatrix}
         +1 & 0 & 0 & 0 & 0 \\
          0 & +1 & 0 & 0 & 0 \\
          0 & 0 & -1 & 0 & 0 \\
          0 & 0 & 0 & -1 & 0 \\
          0 & 0 & 0 & 0 & -1 \\
        \end{pmatrix}.
\end{equation}

This effectivelly breaks $SU(5)\to SM$, after integrating out the ED, at low energies. Furthermore, if the Higgses live in the bulk, only the doublets remain light after orbifolding. Thus we are free of the doublet-triplet splitting problem.

\subsection{$S_4$ from orbifolding}
\label{sec:s4orbi}

For a better geometric display, and following \cite{Altarelli:2006kg,Adulpravitchai:2009id,Adulpravitchai:2010na}, we may redefine $2\pi R_1\Rightarrow 2$ and $2\pi R_2\Rightarrow 1$. We also define $z=x_5+ix_6$. Everything can be easily rescaled to the actual size. Choosing $\theta=\pi/3$, and defining $\gamma=e^{i\pi/3}$, the symmetries of the orbifold from eqs. \ref{eq:tras},\ref{eq:par} become
\begin{equation}
T_1:\ z\to z+2,\ \ \ T_2:\ z\to z+\gamma,\ \ \ Z:z\to -z, \ \ \ Z^{SM}:z-1\to -z+1,
\end{equation}
where the orbifolding symmetry $Z$ leaves four invariant points (actually 4D branes)
\begin{equation}
[z_1,z_2,z_3,z_4]=[0,\ 1,\ \gamma/2,\ 1+\gamma/2],
\end{equation}
that are the branes where we locate some of the fields.
These fixed branes and are permuted by the operations
\begin{equation}
S_1:z\to z+1,\ \ \ S_2:z+\gamma/2,\ \ \ R:z\to\gamma^2 z,\ \ \ P:z\to z^*,\ \ \ P':z\to-z^*,
\end{equation}
which, after orbifolding, generate the remnant symmetry. We can write these operations explicitly $S_1[(12)(34)],\ S_2[(13)(24)],\ R[(243)(1)],\ P[(34)(1)(2)],\ P'[(34)(1)(2)].$
There are only 3 independent transformations since $S_2=R^2\cdot S_1\cdot R,\ \ P=P'$. 

\begin{figure}[h!]
	\centering
	\begin{subfigure}{0.6\textwidth}
		\centering
		\includegraphics[scale=0.6]{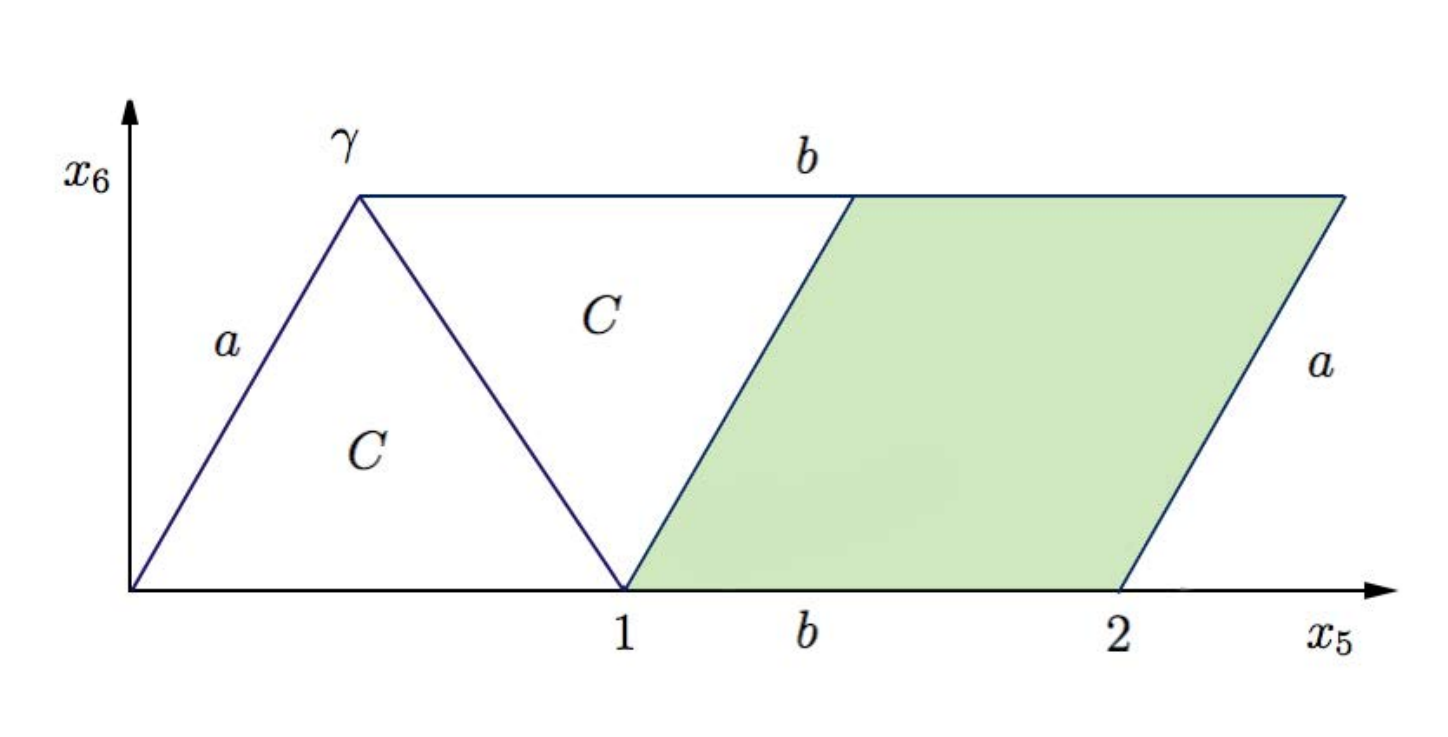}
		\caption{The exta dimensional space. Identifying together sides $a,b$ we obtain $T^2$. The $\mathbb{Z}_2^{SM}$ orbifolding identifies the shaded area with the non shaded. The orbifolding $\mathbb{Z}_2$ identifies both areas labeled $C$.}
	\end{subfigure}%
	\ \ \ \ \
	\begin{subfigure}{0.34\textwidth}
		\centering
		\includegraphics[scale=0.65]{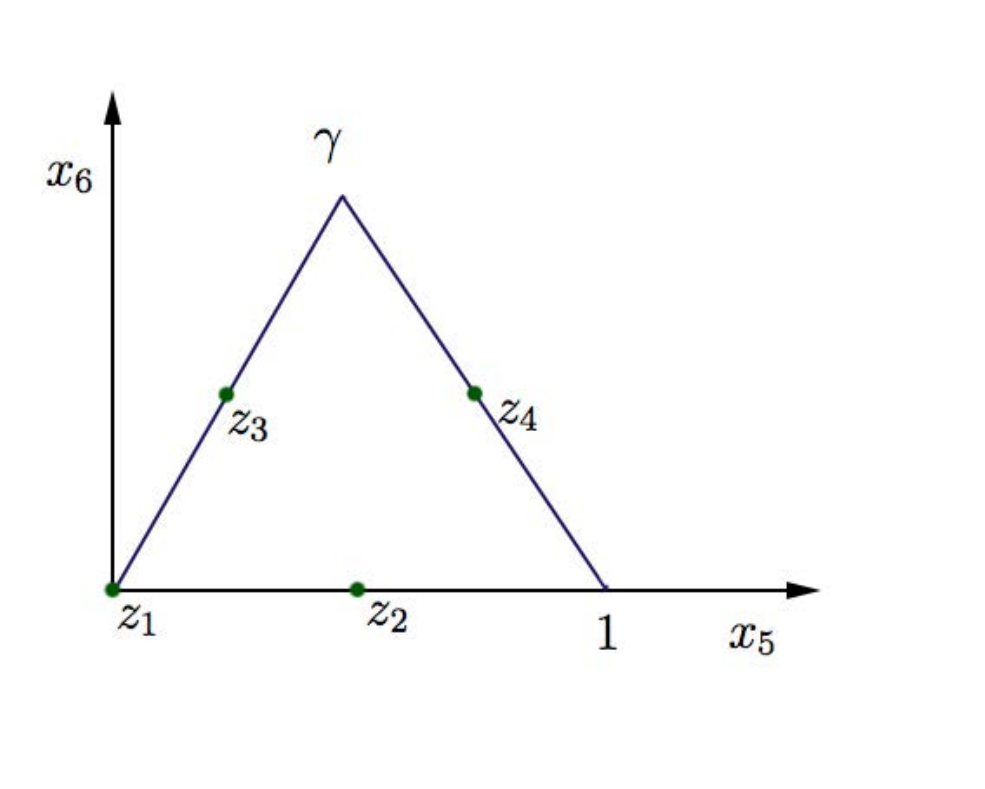}
		\caption{The effective extra dimensional space $T^2/(\mathbb{Z}_2\times \mathbb{Z}_2^{SM})$. This is the whole bulk. The four invariant branes $z_{1,2,3,4}$ are shown.}
	\end{subfigure}
	\begin{subfigure}{0.45\textwidth}
		\centering
		\includegraphics[scale=0.7]{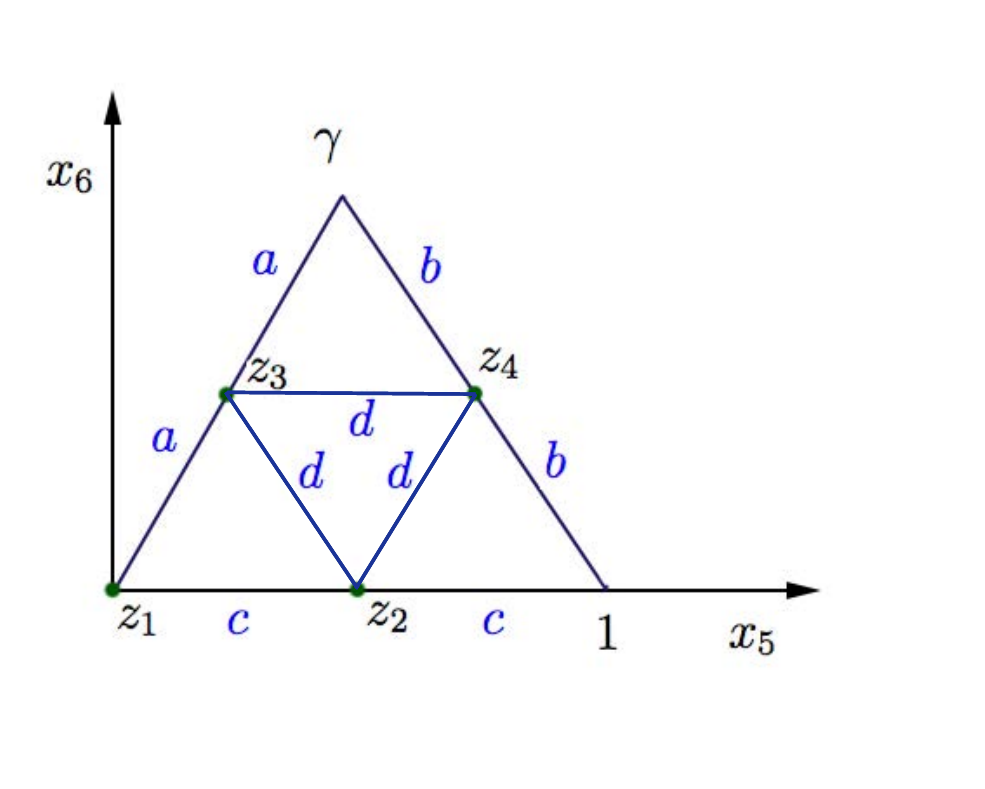}
		\caption{The four branes are permuted by the symmetries $S_1, S_2, R$. These symmetries identify the sides $a,b,c$ while $R$ rotates everything by identifying sides $d$.}
	\end{subfigure}%
	\ \ \ \ \
	\begin{subfigure}{0.45\textwidth}
		\centering
		\includegraphics[scale=0.4]{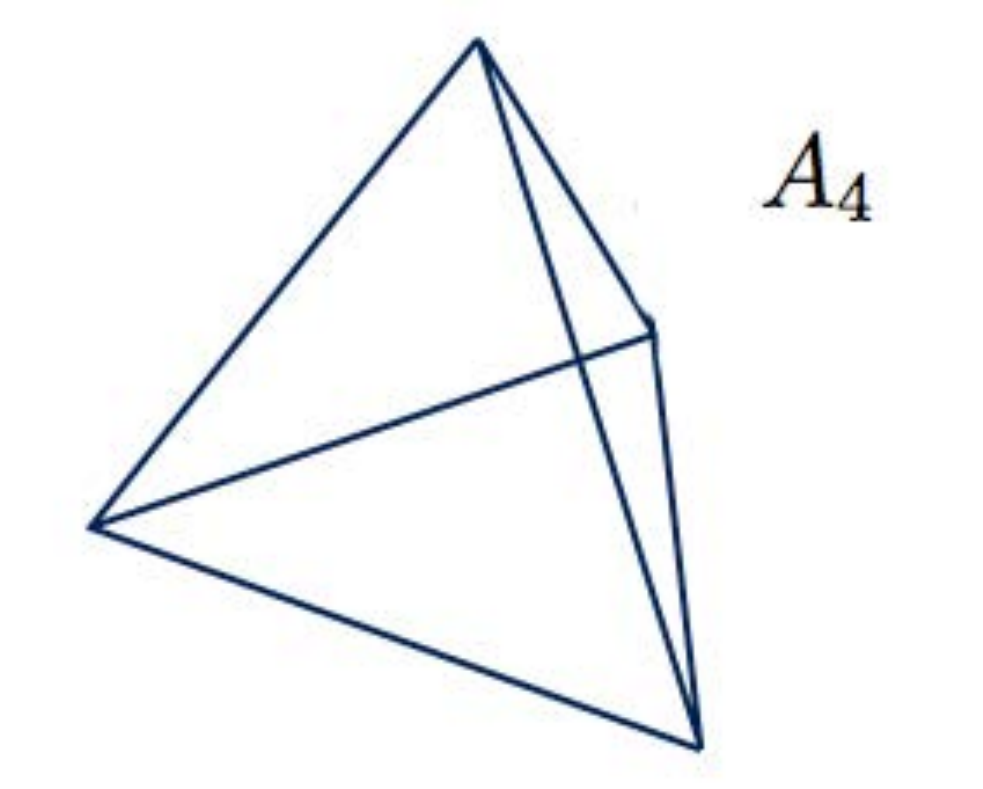}
		\caption{By actually gluing together sides $a,b,c$ we obtain a tetrahedron, whose vertices are related by the symmetry group $A_4$.}
	\end{subfigure}
	\begin{subfigure}{0.45\textwidth}
		\centering
		\includegraphics[scale=0.7]{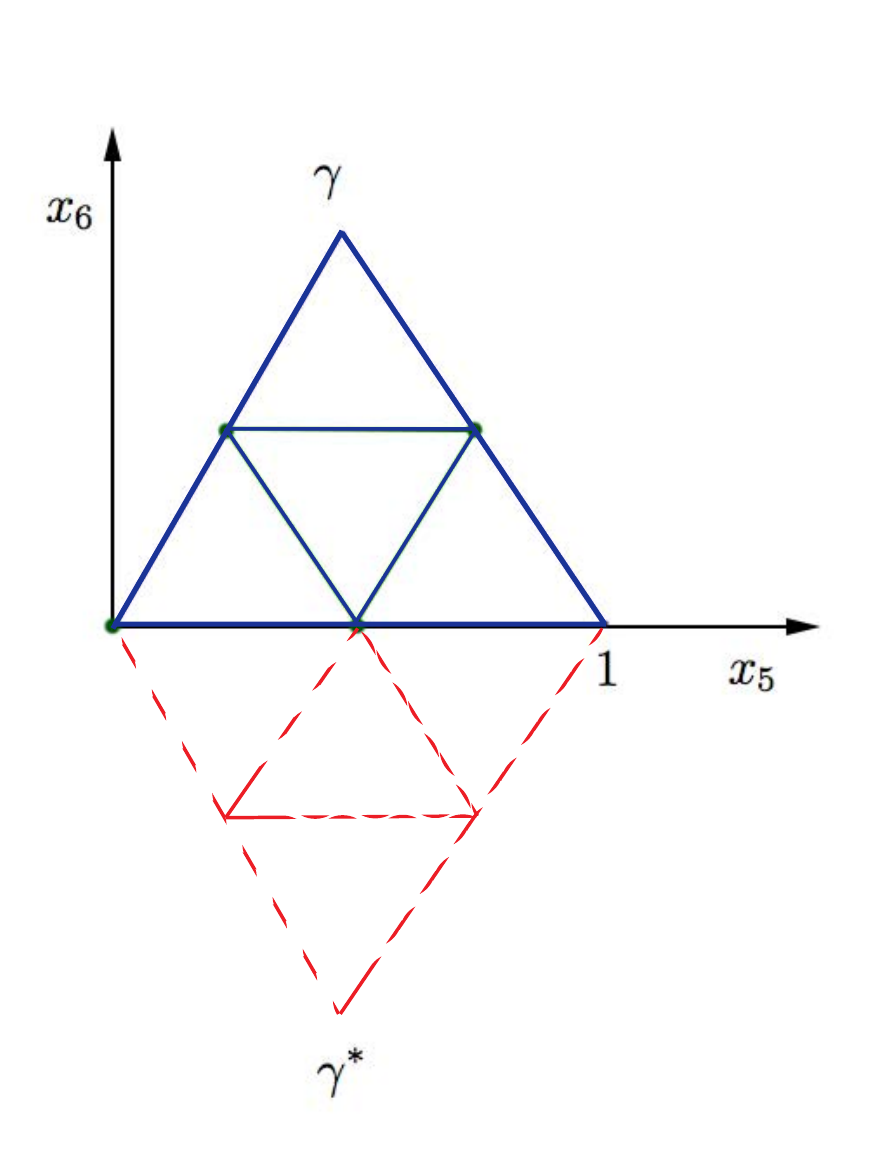}
		\caption{The symmetries $S_1, S_2, R$ generate $A_4$. By also considering independent parities $P,P'$ we obtain the reflected bulk space. }
	\end{subfigure}%
	\ \ \ \ \
	\begin{subfigure}{0.45\textwidth}
		\centering
		\includegraphics[scale=0.4]{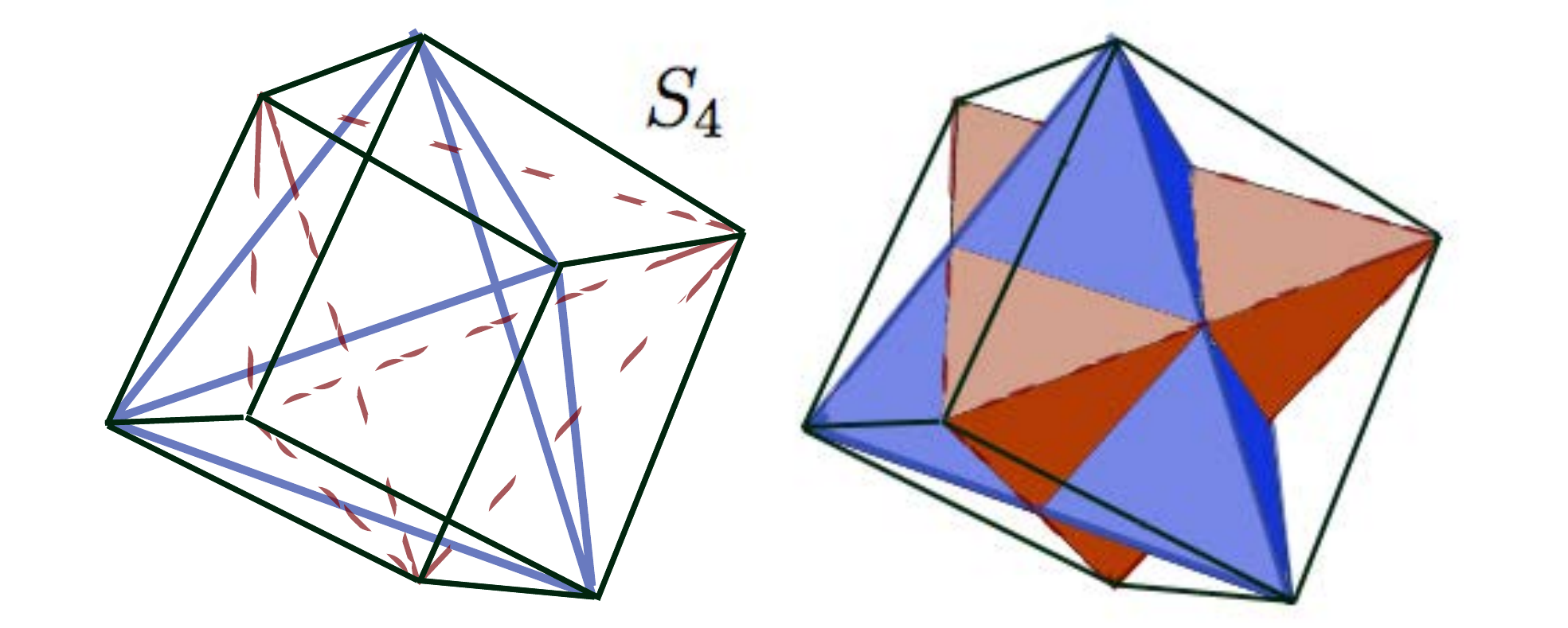}
		\vspace{1cm}
		\caption{Identifying sides $a,b,c$ for each space we obtain a tetrahedron and a reflected one. The pair of tetrahedra lie inside a cube, whose vertices are related by the symmetry group $S_4$. The left image shows all the sides of the tetrahedra while the one on the right is solid for a better visualization.}
	\end{subfigure}
	\caption{Visualization on the remnant $S_4$ symmetry after orbifolding of the extra dimensions.}
	\label{fig:s4orb}
\end{figure}

These symmetry transformations relate to the $S_4$ generators with $S=S_1,\ T=R,\ U=P$ satisfying
\begin{equation}
S^2=T^3=(ST)^3=U^2=(SU)^2=(TU)^2=(STU)^4=1,
\end{equation}
which is the presentation rules for the $S_4$ symmetry \cite{King:2013eh}. Even though only two generators are enough for $S_4$ \cite{Ishimori:2010au}, we prefer this presentation since it shows explicitly its relation to $A_4$. The transformations $S,T$ alone generate $A_4$ \cite{Altarelli:2006kg}. Ignoring the individual parity transformations $P$, the orbifold would have a remnant symmetry of $A_4$. Note that we have not added this symmetry by hand but a remnant of the orbifolding symmetry after compactification. Figure \ref{fig:s4orb} shows a visualization of the remnant $S_4$ symmetry of the extra dimensions after orbifolding.

If we locate a field in each of the branes, they would be transformed between them forming a reducible 4 dimensional representation. We need to obtain the decomposition into irreducible representations \cite{Bazzocchi:2009pv}.
Choosing $S=S_1,\ T=R,\ U=P,$  we obtain the matrices
\begin{equation}
S=\left(\begin{array}{cccc}0&1&0&0
\\ 1&0&0&0
\\ 0 &0&0&1
\\ 0&0&1&0\end{array}\right), \ \ \ 
T=\left(\begin{array}{cccc}1&0&0&0
\\ 0&0&1&0
\\ 0&0&0&1
\\ 0&1&0&0\end{array}\right), \ \ \ 
U=\left(\begin{array}{cccc}1&0&0&0
\\ 0&1&0&0
\\ 0&0&0&1
\\ 0&0&1&0\end{array}\right).
\end{equation}
With the unitary transformation
\begin{equation}
V=\left(\begin{array}{cccc}
-\frac{\sqrt{3}\omega^2}{2}&0&0&\frac{1}{2}
\\ \frac{\omega^2}{2\sqrt{3}}&\frac{\omega^2}{\sqrt{3}}&\frac{\omega^2}{\sqrt{3}}&\frac{1}{2}
\\ \frac{\omega^2}{2\sqrt{3}} &\frac{\omega}{\sqrt{3}}&\frac{1}{\sqrt{3}}&\frac{1}{2}
\\ \frac{\omega^2}{2\sqrt{3}}&\frac{1}{\sqrt{3}}&\frac{\omega}{\sqrt{3}}&\frac{1}{2}\end{array}\right), 
\end{equation}
we can obtain
\begin{equation}
S\to V^\dagger S V=\left(\begin{array}{cc} S_{3}&0 \\ 0&1\end{array}\right),\ \ \ T\to V^\dagger T V=\left(\begin{array}{cc} T_{3}&0 \\ 0&1\end{array}\right), \ \ \ U\to V^\dagger U V=\left(\begin{array}{cc} U_{3}&0 \\ 0&1\end{array}\right),
\end{equation}
so that, the 4 dimensional representation inherited from the branes can be decomposed $\textbf{4}\to \textbf{3}+\textbf{1}$.

If instead we choose $S=-S_1,\ T=R,\ U=P$, the same unitary transformation would decompose $\textbf{4}\to \textbf{3'}+\textbf{1'}$.  It is also possible, with a different choice of generators, to decompose it as $\textbf{4}\to \textbf{2}+\textbf{1}+\textbf{1}$ \cite{Adulpravitchai:2010na}. As only one of these embeddings of $S_4$ can be realized in the model, we choose $S=-S_1,\ T=R,\ U=P$ so that we can only have $\textbf{3'}s$ and $\textbf{1'}s$ on the brane.

A field located on the brane $A_K(x)\delta(z-z_K)$ that transforms as a $\textbf{3'}$ would be written as
\begin{equation}
A_K(x)=V^\dagger_{iK} A_i^{3'}(x).
\end{equation}
A field $A_K(x)\delta(z-z_K)$ that transforms as a $\textbf{1'}$ would be written as
\begin{equation}
A_{K}(x)=V^\dagger_{4K} A^{1'}(x).
\end{equation}

Fields located in the bulk $B(x,z)$ can transform under any irreducible representation with
\begin{equation}\begin{split}
\mathcal{S}&:\ B(x,z)\to S\ B(x,z+1/2),\\
\mathcal{T}&: \ B(x,z)\to T\ B(x,\gamma^2z),\\
\mathcal{U}&:\ B(x,z)\to U\ B(x,z^*),\\
\end{split}\end{equation}
where we use the $S_4$ basis 
\begin{equation}
\begin{array}{c|ccc}
S_4  & S & T & U \\ \hline
{\bf 1,1'}  & 1&1 &\pm 1 \\[2mm]
{\bf 2}  & 
\begin{pmatrix} 1 & 0 \\ 0&1 \end{pmatrix} 
&\begin{pmatrix} \omega & 0 \\ 0&\omega^2 \end{pmatrix}  
& \begin{pmatrix} 0& 1 \\ 1&0 \end{pmatrix} \\[4mm]
{\bf 3,3'} & 
\frac{1}{3}\begin{pmatrix} -1 & 2&2 \\ 2&-1&2 \\2&2&-1 \end{pmatrix} 
&\begin{pmatrix} 1&0&0\\ 0&\omega^2 &0 \\ 0&0&\omega \end{pmatrix}  
& \mp \begin{pmatrix} 1&0&0\\0&0& 1 \\ 0&1&0 \end{pmatrix} 
\\
\end{array}
\label{eq:s4gen}
\end{equation}
that has real $S,U$ and simplify obtaining the CSD3 alignment \cite{King:2011zj}.

We can construct bilinears with one field in the bulk and one in the brane \cite{Burrows:2009pi}. Specifically the singlet $J$ coming from one $\textbf{3'}$ $B(x,z)_i$ located in the bulk and one $\textbf{3'}$ $A_K(x)$ located in each brane would be
\begin{equation}
J=\sum_{iK} B_i(x,z) V^\dagger_{iK} A_K(x)\delta(z-z_K).
\label{eq:6d3}
\end{equation}

After compactification we can treat the flavour symmetry $S_4$ as usual.

\section{Flavon alignment}
\label{sec:flaval}

The predictivity in flavour models comes from the specific flavon structure that define the fermion mass matrices. These alignment is usually fixed by a superpotential. We obtain the alignment through a combination of orbifolding and a superpotential \cite{Kobayashi:2008ih,Adulpravitchai:2010na,Burrows:2010wz}.

We assume that the flavons obtain a VEV through radiative symmetry breaking \cite{Ibanez:1982fr}. There are six flavon multiplets to be aligned. The flavons $\phi_{s,\tau}$ are located in the brane while $\phi_{a,\mu,e}, \rho$ propagate in the bulk and thus are subject to the orbifold boundary conditions.  Since their VEVS are constant, a condition on the boundary implies a condition on the VEV.

We choose the orbifolding parity condition 
\begin{equation}
P'_{SM}=P_{SM}\otimes U,
\end{equation}
where $U$ is one of the $S_4$ generators listed in eq. \ref{eq:s4gen}. The flavons in the brane are not affected by this condition. The flavons in the bulk are not affected by the $P_{SM}$ matrix, since they are GUT singlets. The bulk flavon VEVs must be eigenvectors of the $U$ matrix.

The flavon $\phi_a$ is a $\textbf{3'}$ and it has a negative parity in the boundary condition so that it must comply with
\begin{equation}
\braket{\phi_a}=-U\braket{\phi_a}=-\begin{pmatrix} 1&0&0\\0&0& 1 \\ 0&1&0 \end{pmatrix} \braket{\phi_a}\ \ \ \to \ \ \ \braket{\phi_a}\sim \left(\begin{array}{c}0\\1\\-1\end{array}\right).
\end{equation}

The flavon $\rho$ is a $\textbf{2}$ with positive parity so that it must comply with
\begin{equation}
\braket{\rho}=U\braket{\rho}=\begin{pmatrix} 0& 1 \\ 1&0 \end{pmatrix} \braket{\rho}\ \ \ \to \ \ \ \braket{\rho}\sim  \left(\begin{array}{c}1\\1\end{array}\right).
\end{equation}

The flavon $\phi_e$ is located in the bulk with positive parity, it must comply with
\begin{equation}
\braket{\phi_e}=-U\braket{\phi_e}=\begin{pmatrix} 1&0&0\\0&0& 1 \\ 0&1&0 \end{pmatrix} \braket{\phi_e}\ \ \ \to \ \ \ \braket{\phi_e}\sim \left(\begin{array}{c}a\\b\\b\end{array}\right),
\label{eq:phieor}
\end{equation}
with arbitrary $a,b$. 

To fix the alignment of the flavons in the brane, we make use of a superpotential and the alignment fields $A$ in table \ref{tab:funfields}. The 6d superpotential of brane and bulk fields at leading order is in eq. \ref{eq:6dalign}. After compactification we obtain the simpler looking superpotential
\begin{equation}
\mathcal{W}_A\sim A_1(\phi_\tau)^2+A_2(\phi_e)^2+A_1'(\phi_\mu\phi_\mu+\phi_e\phi_\tau)+A_{3}(\phi_a\phi_\tau-\rho \phi_s),
\label{eq:alpot}
\end{equation}
where we ignore the effective dimensionless constants

The F-term equation coming from $A_1$ fixes
\begin{equation}
\braket{\phi_{\tau}}\sim \left(\begin{array}{c}0\\1\\0\end{array}\right),\left(\begin{array}{c}0\\0\\1\end{array}\right),\left(\begin{array}{c}2\\2x\\-1/x\end{array}\right),
\label{eq:alsin}
\end{equation}
with arbitrary $x$. We choose $\braket{\phi_\tau}$ to be the first solution.

The F-term equation from $A_2$ fixes the VEV
\begin{equation}
\braket{\phi_e}\sim \left(\begin{array}{c}1\\0\\0\end{array}\right),\left(\begin{array}{c}1\\-2\omega^n\\-2\omega^{2n}\end{array}\right),
\end{equation}
so that, together with the orbifold condition from eq. \ref{eq:phieor}, the $\braket{\phi_e}$ is fixed to be the first choice.

The $A_{1}'$ alignment field involves two terms. The contraction $\phi_e \phi_\tau$ enters into the equation and it is is exactly zero since the alignments are fixed to be orthogonal, as show above. So that the F term equation from $A_1'$ fixes 
\begin{equation}
\braket{\phi_{\mu}}\sim \left(\begin{array}{c}0\\1\\0\end{array}\right),\left(\begin{array}{c}0\\0\\1\end{array}\right),\left(\begin{array}{c}2\\2x\\-1/x\end{array}\right),
\label{eq:alsinp}
\end{equation}
where we choose the second solution.

Finally, the F term equation from $A_{3}$ fixes \cite{King:2016yvg}
\begin{equation}
\braket{\phi_s}\sim \left(\begin{array}{c}1\\3\\-1\end{array}\right).
\end{equation}

We remark that $\braket{\phi_{a,s}},\braket{\rho}$ preserve $SU$ while $\braket{\phi_e}, \omega\braket{\phi_\mu}, \omega^2 \braket{\phi_\tau}$ preserve $T$, where $S,T,U$ are the $S_4$ generators.

We have obtained the flavon VEV alignments
\begin{equation}\begin{split}
\braket{\phi_s}=v_s\left(\begin{array}{c}1\\3\\-1\end{array}\right),\ \ \braket{\phi_a}&=v_a\left(\begin{array}{c}0\\1\\-1\end{array}\right),\ \  \braket{\rho}=v_\rho\left(\begin{array}{c}1\\1\end{array}\right),\\
\braket{\phi_e}=v_e\left(\begin{array}{c}1\\0\\0\end{array}\right),\ \ \braket{\phi_\mu}&=v_\mu\left(\begin{array}{c}0\\0\\1\end{array}\right), \ \ 
\braket{\phi_\tau}=v_\tau\left(\begin{array}{c}0\\1\\0\end{array}\right).
\end{split}\end{equation}

We have achieved the so called CSD3 alignment from the flavons $\braket{\phi_{s,a}}$ using orbifolding and the superpotential in eq. \ref{eq:alpot} which is remarkably simple compared to previous ways to achieve it  \cite{King:2016yvg,Bjorkeroth:2015uou,King:2013iva,Bjorkeroth:2017ybg}.

\section{SM fermion mass structure}
\label{sec:smfer}

The Yukawa superpotential originally is 6d and is stated in  eq. \ref{eq:yuk6d}. We will work with the compactified superpotential, assuming that the cutoff scale $\Lambda$ is close enough to the compactification scale , we may write the effective superpotential
\begin{equation}\begin{split}
\mathcal{W}_Y&=y_{ij}^u  H_5 T_i^- T_j^+\left(\frac{\xi}{\Lambda}\right)^{6-i-j}\\
&\quad+y_{33}^\pm H_{\overbar{5}}F\phi_\tau T_3^\pm\frac{1}{\Lambda}+y_{22}^\pm H_{\overbar{5}}F\phi_\mu T_2^\pm\frac{1}{\Lambda}+y_{11}^\pm H_{\overbar{5}}F\phi_eT_1^\pm\frac{1}{\Lambda}
\\ &\quad +y_{23}^\pm H_{\overbar{5}}F\phi_\tau T_2^\pm\frac{\xi}{\Lambda^2}
+y_{13}^\pm H_{\overbar{5}}F\phi_\tau T_1^\pm\frac{\xi^2}{\Lambda^3}+y_{12}^\pm H_{\overbar{5}}F\phi_\mu T_1^\pm\frac{\xi}{\Lambda^2}\\
&\quad+y_a^\nu H_5F\phi_aN_a^c\frac{\xi}{\Lambda^2}+y_s^\nu H_5F\phi_sN_s^c\frac{1}{\Lambda}+y^N_s \frac{\xi^4}{\Lambda^3} N_a^cN_a^c+y^N_s\xi N_s^cN_s^c\\
&\quad+ y_H\frac{\xi^{10}}{\Lambda^9}H_5 H_{\overbar{5}},
\label{eq:yuk}
\end{split}\end{equation}
where the effective dimensionless coupling constants $y$ are expected to be $O(1)$ and real due to the imposed trivial CP symmetry.

The first line in eq. \ref{eq:yuk} gives masses to the up quarks. Since $Q$ comes from $T^-$, while $u^c$ comes from $T^+$, the up quark matrix is not symmetric, as in usual $SU(5)$ theories. The top mass is effectively renormalisable while the others are not.
Defining 
\begin{equation}
\braket{\xi,v_i}/\Lambda=\tilde{\xi},\tilde{v_i},
\label{scaled}
\end{equation}
where $i=e,\mu, \tau, a, s$,
we write the up quark mass matrix \footnote{All the mass matrices are given in the LR convention.}
\begin{equation}
M^u=v_u\left(\begin{array}{ccc}
y_{11}\tilde{\xi}^4 & y_{12}\tilde{\xi}^3 & y_{13}\tilde{\xi}^2\\
y_{21}\tilde{\xi}^3 & y_{22}\tilde{\xi}^2 & y_{23}\tilde{\xi}\\
y_{31}\tilde{\xi}^2 & y_{32}\tilde{\xi} & y_{33}
\end{array}\right).
\end{equation}

The second and third lines of eq. \ref{eq:yuk} give masses to down quarks and charged leptons. The down quark matrix is
\begin{equation}
M^d=v_d\left(\begin{array}{ccc}
y_{11}^-\tilde{v}_e & y_{12}^-\tilde{v}_\mu\tilde{\xi} & y_{13}^-\tilde{v}_\tau\tilde{\xi}^2\\
0 & y_{22}^-\tilde{v}_\mu & y_{23}^-\tilde{v}_\tau\tilde{\xi}\\
0 & 0 & y_{33}^-\tilde{v}_\tau
\end{array}\right),
\end{equation}
while the charged lepton mass matrix is
\begin{equation}
(M^e)^*=v_d\left(\begin{array}{ccc}
y_{11}^+\tilde{v}_e & 0&0 \\
y_{12}^+\tilde{v}_\mu\tilde{\xi}  & y_{22}^+\tilde{v}_\mu & 0\\
y_{13}^+\tilde{v}_\tau\tilde{\xi}^2 & y_{23}^+\tilde{v}_\tau\tilde{\xi} & y_{33}^+\tilde{v}_\tau
\end{array}\right).
\end{equation}
Since $e^c$ comes from $T^+$ and $Q$ comes from $T^-$ the Yukawa terms have different and independent couplings $y^\pm_{ij}$ for each one. This way the charged lepton mass matrix is completely independent of the down quark mass matrix.

The fourth line in eq. \ref{eq:yuk} gives the Dirac neutrino mass matrix and the right handed neutrino Majorana mass matrix
\begin{equation}
M^\nu_D=v_u\left(\begin{array}{cc} 0 & y_s^\nu \tilde{v}_s\\
-y_a^\nu \tilde{v}_a\tilde{\xi}& -y_s^\nu \tilde{v}_s\\
y_a^\nu \tilde{v}_a\tilde{\xi} & 3y_s^\nu \tilde{v}_s
\end{array}\right), \ \ \ 
M^N =\left(\begin{array}{cc}
y^N_a\tilde{\xi}^3&0 \\ 0 &y^N_s
\end{array}\right)\braket{\xi}.
\end{equation}

The RHN are very heavy so that the left handed neutrinos become very light after the seesaw mechanism has been implemented,
\begin{equation}\begin{split}
M^\nu &=M^\nu_D(M^N)^{-1}(M^\nu)^T\\
&=\frac{v_u^2}{\braket{\xi}}\frac{(y^\nu_a)^2\tilde{v}_a^2}{y^N_a\tilde{\xi}}\left(\begin{array}{ccc} 0 &0 &0 \\ 0 & 1 & -1 \\ 0 & -1 &1\end{array}\right)+\frac{v_u^2}{\braket{\xi}}\frac{(y^\nu_s)^2\tilde{v}_s^2}{y^N_s}\left(\begin{array}{ccc} 1 &-1&3 \\ -1 & 1 & -3 \\ 3 & -3 &9\end{array}\right).
\end{split}\end{equation}
This structure for the neutrino mass matrix is called the Littlest Seesaw \cite{King:2013iva}.

\subsection{Low energy parameters and physical phases}

We assume that the scaled VEVs in Eq.\ref{scaled} take hierarchical values,
\begin{equation}
|\tilde{v}_{e}|\ll |\tilde{v}_{\mu}| \ll  |\tilde{v}_{\tau}|, |\tilde{v}_{a}|, |\tilde{v}_{s}|, |\tilde{\xi}|< 1,
\end{equation}
perhaps due to radiative breaking at different scales, The idea is that powers of these VEVs
are responsible for the hierarchies between the fermion masses, allowing all the $y$ parameters 
appearing in the mass matrices of the previous section to be $O(1)$.
However in practice, a couple of these $y$ parameters will need to be of order $5\%$.
To the extent that these parameters are $O(1)$, 
our model may be regarded as providing a ``natural'' explanation of the quark and lepton (including neutrino) masses
and mixings, including the CP phases, as we now discuss.

We have imposed trivial CP symmetry, so that all coupling constants $y$ are real. However, the same mechanism that drives the flavon VEVs may spontaneously break CP. We will assume this is the case by having all flavon VEVs generally complex, with phases
\begin{equation}
\arg(v_{f})=\eta_f,
\end{equation}
where $f$ is each flavon. We can always absorb phases into the fermion fields and we redefine
\begin{equation}\begin{split}
u_{L,R}^2\to e^{-i\eta_\xi}u_{L,R}^2,\ &\ \  u_{L,R}^1\to e^{-2i\eta_\xi}u_{L,R}^1,\\
d_R^1\to e^{-i\eta_e }d_R^1,\ &\ \ d_R^2\to e^{-i\eta_\mu }d_R^2, \ \ \ d_R^3\to e^{-i\eta_\tau }d_R^3,\\
e_L^1\to e^{i\eta_e }e_L^1,\ &\ \ e_L^2\to e^{i\eta_\mu }e_L^2, \ \ \ e_L^3\to e^{i\eta_\tau }e_L^3,\\
\nu_L^i\to e^{-i\eta_a+i\eta_\xi}\nu_L^i,\ &\ \ \nu_L^2\to -\nu_L^2.
\end{split}\end{equation}
With these phase redefinitions, the charged fermion mass matrices of the previous section may be rewritten in terms of explicitly real 
parameters and physical phases, 
\begin{equation}
\begin{split}
M^u&=v_u\left(\begin{array}{ccc}
y_{11}\ |\tilde{\xi}|^4 & y_{12}\ |\tilde\xi|^3 & y_{13}\ |\tilde\xi|^2 \\
y_{21}\ |\tilde\xi|^3 & y_{22}\ |\tilde\xi|^2 & y_{23}\ |\tilde\xi| \\
y_{31}\ |\tilde\xi|^2  & y_{32}\ |\tilde\xi|   & y_{33}
\end{array}\right)\\
M^d&=v_d\left(\begin{array}{ccc}
y_{11}^-\ |\tilde v_e| & y_{12}^-\ |\tilde v_\mu \tilde\xi| e^{i\eta_\xi} & y_{13}^-\ |\tilde v_\tau \tilde\xi^2| e^{2i\eta_\xi}\\
0 & y_{22}^-\ |\tilde v_\mu|  & y_{23}^-\ |\tilde v_\tau\tilde\xi| e^{i\eta_\xi}\\
0 & 0 & y_{33}^-\ |\tilde v_\tau|
\end{array}\right)\\
M^e&=v_d\left(\begin{array}{ccc}
y_{11}^+\ |\tilde v_e|   & 0&0 \\
y_{12}^+\ |\tilde v_\mu \tilde\xi|  e^{-i\eta_\xi}  & y_{22}^+\ |\tilde v_\mu| & 0\\
y_{13}^+\ |\tilde v_\tau \tilde\xi^2| e^{-2i\eta_\xi} & y_{23}^+\ |\tilde v_\tau\tilde\xi| e^{-i\eta_\xi} & y_{33}^+\ |\tilde v_\tau|
\end{array}\right),
\label{eq:fitmat}
\end{split}
\end{equation}
while the low energy neutrino mass matrix, after the seesaw mechanism has been implemented, may be expressed as \footnote{We use the convention $-\frac{1}{2}M^\nu \bar{\nu}_L\nu^c_L$ and $-\frac{1}{2}M^N\bar{\nu}^c_R\nu_R$, for Majorana masses.},
\begin{equation}
M^\nu =\mu_a\left(\begin{array}{ccc} 0 &0 &0 \\ 0 & 1 & 1 \\ 0 & 1 &1\end{array}\right)+\mu_s \ |\tilde\xi| e^{i\eta}\left(\begin{array}{ccc} 1 &1&3 \\ 1 & 1 & 3 \\ 3 & 3 &9\end{array}\right),
\label{eq:fitnu}
\end{equation}
where
\begin{equation}
\eta=2\eta_s-2\eta_a+\eta_\xi,
\end{equation}
and
\begin{equation}
\mu_{a,s}=\frac{(v_uy_{a,s}^\nu)^2}{|v_\xi|y^N_{a,s}}.
\end{equation}

So finally we have only 2 physical phases $\eta, \eta_\xi$, plus 2 left handed neutrino mass parameters $\mu_{a,s}$
and 21 dimensionless $O(1)$ parameters $y$,
noting that the VEV ratios $|\tilde v_i|,|\tilde\xi|$ are not physical low energy degrees of freedom but are used just to absorb the hierarchies between fermion masses.

\subsection{Numerical Fit}

With these parameters we may perform a fit for the fermion masses and mixings, comparing the model to these values 
run up to the GUT scale. In table \ref{tab:ferobs} we show the observables that can be obtained using the parameters in table \ref{tab:parameters}. We can fit perfectly the full SM fermion content observables run up to the GUT scale with $\chi^2\approx 0$, choosing $\tan\beta=5$ and assuming negligible SUSY threshold corrections. We can obtain an equally good fit with different $\tan\beta$ and this one is chosen arbitrarily.

\begin{table}
	\centering
	\footnotesize
	\renewcommand{\arraystretch}{1.1}
	\begin{tabular}{| c |  c|}
\hline
Parameter &  Model \\
\hline
$\theta_{12}^q$ & $13.026^\circ$  \\
$\theta_{13}^q$ & $0.193^\circ$ \\
$\theta_{23}^q$ & $2.237^\circ$\\
$\delta_q$ & $69.21^\circ$\\
\rule{0pt}{4ex}%
$y_u$  & $2.92\times 10^{-6}$ \\
$y_c$  & $1.43\times 10^{-3}$  \\
$y_t$  & $5.34\times 10^{-1}$ \\
\rule{0pt}{4ex}%
$y_d$  & $4.81\times 10^{-6}$ \\
$y_s$  & $9.52\times 10^{-5}$  \\
$y_b$  & $5.38\times 10^{-3}$  \\
\hline
\end{tabular}
	\hspace*{0.5cm}
	\begin{tabular}{| c | c|}
\hline
Parameter   & Model \\
\hline
$\theta_{12}^l$ & $33.63^\circ$ \\
$\theta_{13}^l$ & $8.54^\circ$  \\
$\theta_{23}^l$ & $47.2^\circ$ \\
$\delta_l$ & $234.15^\circ$\\
\rule{0pt}{4ex}%
$y_e$  & $1.97\times 10^{-6}$ \\
$y_\mu$  & $4.16\times 10^{-4}$  \\
$y_\tau$  & $7.07\times 10^{-3}$ \\
\rule{0pt}{4ex}%
$\Delta m_{21}^2/eV^2$  & $7.51\times 10^{-5}$ \\
$\Delta m_{31}^2/eV^2$  & $2.52\times 10^{-3}$  \\
\rule{0pt}{4ex}%
$m_1/meV$  & $0$\\
$m_2/meV$  &$8.67$\\
$m_3/meV$  & $50.23$\\
\rule{0pt}{4ex}%
$\alpha_{23}$ & $33.85^\circ$ \\
\hline
\end{tabular}
	\caption{Fermion masses and mixings fitted in the model. They resemble exactly the observed ones with $\chi^2\approx 0$. The observables are at the GUT scale and with $\tan\beta=5$. The quark masses, charged lepton masses and CKM parameters come from \cite{Antusch:2013jca}. The neutrino observables come from \cite{Esteban:2016qun}. The fit has been performed using the Mixing Parameter Tools (MPT)  package~\cite{Antusch:2005gp}.} 
	\label{tab:ferobs}
\end{table}

\begin{table}
	\centering
	\footnotesize
	\renewcommand{\arraystretch}{1.1}
	\begin{tabular}{|c|c|}
	\hline
		Parameter & Value \\ 
		\hline
		$y_{11}$ &0.9 \\
		$y_{12}$ & 1.13 \\
		$y_{13}$ & 1.75 \\
		$y_{21}$ & -0.42 \\
		$y_{22}$ &-0.64 \\
		$y_{23}$ & -0.90 \\
		$y_{31}$ & -1.78 \\
		$y_{32}$ & -0.05 \\
		$y_{33}$ & 0.53 \\
		\hline
	\end{tabular}
	\hspace*{0.5cm}
	\begin{tabular}{|c|c|}
	\hline
		Parameter & Value \\ 
		\hline
		$y_{11}^-$ & 2.45 \\
		$y_{12}^-$ & 0.53\\
		$y_{13}^-$ & 0.37 \\
		$y_{22}^-$ & 1.28 \\
		$y_{23}^-$ & -2.53 \\
		$y_{33}^-$ & 0.05 \\
		\rule{0pt}{4ex}%
		$y_{11}^+$ & -1.00 \\
		$y_{12}^+$ & 0.58 \\
		$y_{13}^+$ & -0.59 \\
		$y_{22}^+$ & 2.12 \\
		$y_{23}^+$ & 0.21 \\
		$y_{33}^+$ & -0.17 \\
		\hline
	\end{tabular}
	\hspace*{0.5cm}
	\begin{tabular}{|c|c|}
	\hline
		Parameter & Value \\ 
		\hline
		$\mu_a$ & $26.64\ meV$\\
		$\mu_s$ & $53.74\ meV$\\
		\rule{0pt}{4ex}%
		$\eta_\xi$ & $-8\pi/9$\\
		$\eta$ & $-2\pi/3$\\
		\hline
		VEV ratio & Value \\ 
		\hline
		$|\tilde\xi|$ & 0.05\\
		$|\tilde v_\tau|$ & 0.2\\
		$|\tilde v_\mu|$ & 0.001\\
		$|\tilde v_e|$ & 0.00001\\
		\hline
	\end{tabular}
	\caption{The 25 input parameters that enter into the mass matrices in Eqs.~\ref{eq:fitmat},\ref{eq:fitnu} giving the fit in table \ref{tab:ferobs}.
	We also show the assumed VEV ratios, which are not independent parameters.} 
	\label{tab:parameters}
\end{table}

The VEV ratios $|\tilde v_i|$ in table \ref{tab:parameters} are not physical degrees of freedom and are chosen to generate the hierarchies between the fermion mass parameters. For the given choice of VEV ratios,
most of the dimensionless real parameters $y$ turn out to be $O(1)$, although a couple of these parameters,
$y_{32}$ and $y^-_{33}$, are about 5\%,
and a couple more, $y^+_{23}$ and $y^+_{33}$, are about 20\%. However, this choice of parameters does not
appear to be a statistically significant departure from the $O(1)$ expectation.
Furthermore we remark that the two physical phases come out to be multiples of the ninth root of unity.

Once the VEV ratios are chosen, they generate extra predictions. From the last line of eq. \ref{eq:yuk} we may obtain the $\mu$ term. Knowing $|\tilde\xi|$ and that the compactification scale is at the GUT scale we may obtain
\begin{equation}
\mu\sim \frac{\xi^{10}}{\Lambda^9}\sim 1\ TeV,
\end{equation}
which is the correct scale for the $\mu$ term.

From the last two terms in the forth line of eq. \ref{eq:yuk}, we can also estimate the scale for the RHN Majorana masses, so that
\begin{equation}
M_2\sim \xi\sim 10^{15} GeV,\ \ \ M_1\sim 10^{10} GeV,
\label{eq:rhnm}
\end{equation}
as we shall see below, this has consequences for leptogenesis.

The parameters fix the complete $U_{PMNS}$ matrix, including the Majorana phases. In the PDG parametrisation
\begin{equation}\small
U_{PMNS}=\left(\begin{array}{ccc}
c_{12}c_{14} & s_{12} c_{13} & s_{13} e^{-i\delta}\\
-s_{12}c_{23}-c_{12}s_{23}s_{13} e^{i\delta} &c_{12}c_{23}-s_{12}s_{23}s_{13}e^{i\delta} & s_{23}c_{13}\\
s_{12}s_{23}-c_{12}c_{23}s_{13}e^{i\delta} & -c_{12}s_{23}-s_{12}c_{23}s_{13}e^{i\delta}& c_{23} c_{13}
\end{array}\right)
\left(\begin{array}{ccc}
1&0&0\\
0&e^{i\alpha_{21}/2}&0\\
0&0&e^{i\alpha_{31}/2}
\end{array}\right),
\end{equation}
where, since in our model $m_1=0$, only one of the Majorana phases $\alpha_{23}=\alpha_{21}-\alpha_{31}$ is physical. This phase prediction is listed in table \ref{tab:ferobs}. This phase is determined by the effective mass  \cite{Patrignani:2016xqp}
\begin{equation}
|m_{ee}|=|m_2s_{12}^2c_{13}^2e^{i\alpha_{23}}+m_3 s_{13}^2e^{-2i\delta}|\approx 1.5 \ meV,
\end{equation}
which is observable through neutrinoless double beta decay. The effective mass predicted by our model still lies beyond the reach of current experiments.

\subsection{The $U_{PMNS}$ matrix}
\label{sec:upmns}

This neutrino mass matrix is diagonalized by an unitary matrix
\begin{equation}
diag(m_1^\nu,m_2^\nu,m_3^\nu)=U_\nu M^\nu U_\nu^T.
\end{equation}

The charged lepton matrix is diagonalized by two different unitary matrices $V_{eL},V_{eR}$ such that
\begin{equation}\begin{split}
diag(m_e,m_\mu,m_\tau)&=V_{eL} M^e V_{eR}^\dagger,\\
diag(m_e^2,m_\mu^2,m_\tau^2)&=V_{eL}M^{e}M^{e\dagger}V_{eL}^\dagger=V_{eR}M^{e\dagger} M^e V_{eR}^\dagger.
\end{split}\end{equation}
 
Ignoring phases and using the fit parameters from table \ref{tab:parameters}, we may approximate 
\begin{equation}
\begin{split}
M^e&\approx v_d\left(\begin{array}{ccc}
-1\times10^{-5}  & 0&0 \\
3\times 10^{-4}  & 2\times 10^{-3}  & 0\\
-3\times 10^{-4}  & 2\times 10^{-3} & -3\times 10^{-2}
\end{array}\right),\\
M^e M^{e\dagger} &\approx v_d^2\left(\begin{array}{ccc}
10^{-10}  & -3\times 10^{-10}&3\times 10^{-9} \\
-3\times 10^{-10}  & 4\times 10^{-6}  &4\times 10^{-6} \\
3\times 10^{-9}   &4\times 10^{-6} &  10^{-2} 
\end{array}\right),\\
M^{e\dagger} M^e &\approx v_d^2 \left(\begin{array}{ccc}
9\times 10^{-8}   & -6\times 10^{-7} &1\times 10^{-5}\\
-6\times 10^{-7}  & 9\times 10^{-6}  &-7\times 10^{-5}\\
1\times 10^{-5} &-7\times 10^{-5}&1\times 10^{-3}
\end{array}\right),
\label{eq:anguni}
\end{split}
\end{equation}
from which we obtain the rotation angles of the unitary matrices to be
\begin{equation}\begin{split}
\theta_{12}^L\approx -10^{-4},\hspace{7mm}\theta_{23}^L&\approx 4\times 10^{-4},\hspace{7mm}\theta_{13}^L\approx 10^{-8},\\
\theta_{12}^R\approx -5\times 10^{-2},\hspace{7mm}\theta_{23}^R&\approx -7\times 10^{-2},\hspace{7mm}\theta_{12}^R\approx 9\times 10^{-5}.
\end{split}\end{equation}

The PMNS matrix is 
\begin{equation}
U_{PMNS}=V_{eL}U_\nu^{\dagger},
\end{equation}
and only the $\theta_L$ angles enter into it. Furthermore, from eq. \ref{eq:anguni}, we see that the $\theta_L$ angles are very small and the PMNS matrix becomes approximately
\begin{equation}
U_{PMNS}\approx U_\nu^{\dagger},
\end{equation}
and is dominated by the neutrino sector.
Hence, even though we have 25 parameters in the fit, only 3 of these parameters, namely $\mu_a,\mu_s,\eta$, generate the entire spectrum of 
neutrino masses $m_i$ and lepton mixings and phases $U_{PMNS}$. This way we show that the theory is highly predictive in the neutrino sector, since 3 parameters generate the 3 neutrino masses and the 6 PMNS parameters, hence 6 predictions. For example, it predicts $m_1=0$ with only one physical Majorana phase. The remaining 4 physical predictions are determined from the fit.

The predictive power comes from the CSD3 alignment. We have checked that with a diagonal charged lepton mass matrix, 2 RHN and the CSD3 alignment we would still lead to a pretty good fit to the neutrino masses and $U_{PMNS}$ within one sigma. However, even though the off-diagonal charged lepton corrections from $\theta_L$ are very small, it turns out that they do give small deviations to pure CSD3, and help to obtain the perfect fit
with $\chi^2\approx 0$.

\section{Solving the strong CP problem}
\label{sec:strcp}

The Strong CP problem is that the parameter $\overbar{\theta}$ multiplying the topological gauge term is very close to zero. In the SM, there is no reason to be so. 

When CP symmetry is assumed from the beginning, the topological term is forbidden. However, since the effective Yukawa couplings must be complex, this term is generated again after spontaneous CP breaking. The constraint is \cite{Baker:2006ts}
\begin{equation}
\overbar{\theta}=\theta_{QCD}-\theta_q<10^{-10},
\end{equation}
since we impose trivial CP symmetry, we know that $\theta_{QCD}=0$.
The parameter  
\begin{equation}
\theta_q=\arg \det(Y^u Y^d),
\end{equation}
must be then very small. 

In eq. \ref{eq:fitmat} it is shown that the up quark mass matrix in our model is real so that we have $\theta_q=\arg \det(Y^d)$. Also in eq. \ref{eq:fitmat}, we show that the down quark matrix has a triangular structure so that its determinant is the product of the diagonal, which is real. Thus we have obtained that
$\theta_q=0$ and the Strong CP symmetry is solved

There can be higher dimensional contributions that break the triangular structure of the down quark mass matrix or add complex contributions to the up quark matrix. Due to the $U(1)$ symmetry, all the extra terms have to be proportional to 
\begin{equation}
\frac{\mu \braket{H_5 H_{\overbar{5}}}}{\Lambda^3}\sim 10^{-26},
\end{equation}
so that they are completely negligible.

We have shown thus, that in our model there is no Strong CP problem. This has been solved due to the structure of the quark mass matrices fixed by the $S_4$ symmetry \cite{Spinrath:2015ixa}. This mechanism has already been shown previously \cite{Bjorkeroth:2015ora}. We remark that this is actually different than the Nelson-Barr mechanism since it requires having vector-like quarks \cite{Nelson:1983zb}.

\section{Leptogenesis}
\label{sec:leptg}

The universe is made up mostly of matter with an astounding lack of antimatter. This is called the Baryon Asymmetry of the Universe (BAU). Even thou the SM has everything to explain the BAU, it predicts it to be orders of magnitude smaller than it actually is.

One of the most appealing mechanism to explain the BAU is through Leptogenesis, where there is a CP asymmetric decay of the lightest RHN that generates an asymmetry between leptons and antileptons. This lepton asymmetry is transformed into baryon through non perturbative sphaleron processes \cite{DiBari:2012fz}.

Our model has every ingredient necessary to generate the BAU through leptogenesis. As noted in section \ref{sec:upmns}, charged lepton corrections are negligible. Therefore there is only one CP violating phase $\eta$ in the neutrino sector, making it the leptogenesis phase in the sense that the BAU is proportional to $\sin \eta$ (see \cite{Bjorkeroth:2015tsa} for more details). It also provides a direct link with laboratory CP violating phenomena since the Dirac and Majorana phases appearing in the PMNS matrix are both controlled by the single leptonic input phase $\eta$.

Analytic formulas, giving PMNS parameters including CP phases in terms of input parameters including $\eta$, have been calculated in \cite{King:2015dvf}. In the CSD3 alignment, a good fit for all the neutrino data requires this phase to be very close to $\eta\approx 2\pi/3$, so here we assume this value~\cite{Ballett:2016yod}.

The detailed calculation of the BAU through Leptogenesis when the neutrino mass matrix has the CSD3 alignment has already been done 
\cite{Bjorkeroth:2015tsa}. To generate the correct BAU, the lightest RHN mass has to be
\begin{equation}
M_1\approx 3.9\times 10^{10}\ GeV.
\end{equation}
In eq. \ref{eq:rhnm}, we see that our model predicts $M_1$ to be exactly at this scale so that the BAU can be naturally explained un our model through Leptogenesis, just by fitting one $O(1)$ parameter, $y^N_s$.

\section{Proton decay}
\label{sec:prodec}

In GUTs, proton decay is mediated by the usual Higgs triplets and extra gauge fields. However proton decay has not yet been observed and any model has to comply with the constraint for the proton lifetime \cite{Patrignani:2016xqp}
\begin{equation}
\tau_p>10^{29}\ {\rm yrs}.
\end{equation}

Models based on $SO(10)$ GUT symmetry that breaks $M_{GUT}\approx 2\times 10^{16}\ GeV$ and gives masses to the Higgs triplets at this scale predict the proton decay rate to be $\tau_p\sim 10^{29}-10^{30}\ {\rm yrs} $ \cite{Murayama:1994tc}.
In this model GUT breaking happens due to orbifolding, at $M_{GUT}\sim 2\times 10^{16}\ {\rm GeV}$. Furthermore, doublet-triplet splitting happens also due to orbifolding at this scale, so that the present model complies with these constraints. 

Extra dimensional models have extra contributions to proton decay coming from the KK modes. In our model the orbifold conditions break the GUT symmetry so that we have the compactification scale at the GUT scale $M_c\sim 2\times 10^{16}\ {\rm GeV}$. Since the compactification scale is so large, the orbifold processes that contribute to proton decay are at least 3 orders of magnitude smaller than the usual 4d terms
\cite{Altarelli:2001qj}. Even though these processes would give a nice signature for extra dimensions, in our model they don't enhance proton decay significantly.

There may be extra contributions from other effective terms, involving the extra fields that we have added. However in this model the $U(1)$ symmetry does not allow any extra contributions to proton decay coming from the new fields.

\section{Conclusion}

In this paper we have proposed a 6d model with a SUSY $SU(5)$ gauge symmetry containing 22 superfields,
which is a relatively low number as compared to other flavour models as complete as this.
An $S_4$ Family Symmetry emerges from the orbifold, by a generalisation of the mechanism previously used to obtain $A_4$.
The orbifolding breaks $SU(5)$ and solves the doublet triplet splitting problem in the usual way.

In addition, we have obtained the CSD3 vacuum alignment through $F$-term equations and orbifold boundary conditions. 
This is the most efficient method in the literature for obtaining this alignment, 
leading to the highly predictive Littlest Seesaw structure, with 3 input parameters predicting 9 observables in the neutrino sector,
including almost maximal atmospheric mixing and maximal leptonic CP violation,
which agree perfectly with current neutrino oscillation measurements.

Below the compactification scale, after the GUT and Family Symmetry groups are broken, the effective low energy theory is just the 
MSSM, supplemented by two extra RHNs.
However the model predicts a constrained set of input parameters, for example a $\mu$ term with the correct magnitude can be
achieved. Also the Yukawa matrices in the down/charged lepton sectors
have upper/lower triangular form.

Charged fermion masses are naturally hierarchical due to a hierarchy of flavon VEVs, with dimensionless parameters consistent with
an $O(1)$ expectation,
and there are very small charged lepton mixing contributions to the PMNS matrix due to the lower triangular Yukawa matrix.
This feature preserves the successful predictions of the Littlest Seesaw model, including 
neutrino masses with normal ordering with the lightest neutrino mass being zero.

The model spontaneously breaks CP symmetry, with only a single input phase parameter in each of the lepton and quark sectors,
leading to successful CP violation in the quark and lepton sectors.
Due to the upper triangular form of the down quark Yukawa matrix, the single input quark phase and the 
$U(1)$ controlling higher order operators, we show that the model is able to solve the Strong CP problem. 
It also explains the Baryon Asymmetry of the Universe (BAU) through leptogenesis,
with the leptogenesis phase directly linked to the Dirac and Majorana phases.

\subsection*{Acknowledgements}

SFK acknowledges the STFC Consolidated Grant ST/L000296/1 and the European Union's Horizon 2020 Research and Innovation programme under Marie Sk\l{}odowska-Curie grant agreements Elusives ITN No.\ 674896 and InvisiblesPlus RISE No.\ 690575.

\appendix

\section{6d superpotential}
\label{app:6dpot}

The superpotential that generates the Yukawa terms must be fundamentally 6 dimensional. Every superfield has mass dimension 2, and the superpotential must have dimension 5, therefore every interaction term must be non renormalizable. This is the reason it is rarely use. We will show it for completeness, and will assume a single mass scale $\Lambda$ so that the 4 dimensional superpotential in eq. \ref{eq:yuk} comes from 

\begin{equation}\begin{split}
\mathcal{W}_{Y6d}&\sim   H_5 T_i^- T_j^+\frac{1}{\Lambda}\left(\frac{\xi}{\Lambda^2}\right)^{6-i-j}\\
&\quad+ H_{\overbar{5}}F\phi_\tau T_3^\pm\frac{\delta^2(z-z_b)}{\Lambda^3}+ H_{\overbar{5}}F\phi_\mu T_2^\pm\frac{\delta^2(z-z_b)}{\Lambda^3}+ H_{\overbar{5}}F\phi_eT_1^\pm\frac{\delta^2(z-z_b)}{\Lambda^4}
\\&\quad+ H_{\overbar{5}}F\phi_\tau T_2^\pm\xi\frac{\delta^2(z-z_b)}{\Lambda^5}+ H_{\overbar{5}}F\phi_\tau T_1^\pm\xi^2\frac{\delta^2(z-z_b)}{\Lambda^7}+ H_{\overbar{5}}F\phi_\mu T_1^\pm\xi\frac{\delta^2(z-z_b)}{\Lambda^5}
\\ &\quad+ H_5F\phi_aN_a^c\xi\frac{\delta^2(z-z_b)}{\Lambda^6}+H_5F\phi_sN_s^c\frac{\delta^2(z-z_b)}{\Lambda^3}+\xi^4 N_a^cN_a^c\frac{1}{\Lambda^7}+\xi N_s^cN_s^c\frac{1}{\Lambda}
\\&\quad+\frac{\xi}{\Lambda^{19}}H_5 H_{\overbar{5}},
\label{eq:yuk6d}
\end{split}\end{equation}
where the flavour contractions $3'\times 3'\to 1$ of brane and bulk fields come from bilinears shown in eq. \ref{eq:6d3}.  Eq. \ref{eq:yuk} appears after integrating out the EDs and assuming that the compactification scale is at the same scale $\Lambda$.

In the same way, the alignment superpotential comes from the 6 dimensional
\begin{equation}
\mathcal{W}_{A6d}\sim A_1(\phi_\tau)^2\frac{\delta^2(z-z_b)}{\Lambda}+A_2(\phi_e)^2\frac{1}{\Lambda}+A_1'\left(\phi_\mu\phi_\mu+\phi_e\phi_\tau\frac{1}{\Lambda}\right)+A_{3}(\phi_a\phi_\tau-\rho \phi_s)\frac{\delta^2(z-z_b)}{\Lambda},
\label{eq:6dalign}
\end{equation}
where, again we have simplified the contractions shown in eq. \ref{eq:6d3}.

\section{Discrete shaping symmetries}
\label{app:discrete}

The simplest and most effective shaping symmetry to obtain our current model is with the $U(1)$ symmetry we have used. However the shaping symmetry is usually taken to be a discrete one. In this section we show how to build the model with an alternate discrete $\mathbb{Z}_5\times \mathbb{Z}_5\times \mathbb{Z}_2 \times \mathbb{Z}_2$. The charge assignment for each field is listed in table \ref{tab:funfield55}. This effectively generates the same model with a few small differences.

\begin{table}
\centering
\footnotesize
\captionsetup{width=0.9\textwidth}
\begin{tabular}{| c | c@{\hskip 5pt}c |  c c c c c|}
\hline
\multirow{2}{*}{\rule{0pt}{4ex}Field}	& \multicolumn{7}{c |}{Representation} \\
\cline{2-8}
\rule{0pt}{3ex}			& $S_4$ & $SU(5)$ &  $\mathbb{Z}_5$ &$\mathbb{Z}_5$ & $\mathbb{Z}_2$ & $\mathbb{Z}_2$ & $P_{SM}'$ \\ [0.75ex]
\hline \hline
\rule{0pt}{3ex}%
$F$ 			& 3' & $\bar{5}$  &  0 & 4& 0&  0& Brane\\
$T_1^\pm$ 			& 1& $10$ & 3 & 0 & 1 & 0& $\pm 1$\\
$T_2^\pm$ 			& 1& $10$ & 4 & 0 &  0& 0& $\pm 1$\\
$T_3^\pm$ 			& 1& $10$  & 0 & 0 & 1&  0 & $\pm 1$\\
$N_s^c$ 			& 1& $1$  & 2 & 0 & 1& 1 & +1\\
$N_a^c$ 			& 1& $1$ & 3 & 0 & 1& 1 &+1\\
$H_5$ 			& 1& $5$  & 0 & 0 & 0 & 0 &$+1$\\
$H_{\bar{5}}$ 			& 1& $\bar{5}$ & 2 & 0 & 0&  0 &$+1$\\
\rule{0pt}{3ex}%
$\xi$ 			& 1& 1 & 1& 0 & 1& 0 & +1\\
$\rho$ 			& $2$& $1$ & 1 & 1 & 1& 0 &$+1$\\
$\phi_s$ 			& 3'& 1  & 3 & 1 & 1&  1 &Brane \\
$\phi_a$ 			& 3'& 1  & 1 & 1& 1 & 1& $-1$\\
$\phi_\tau$ 			& 3'& 1 & 3 & 1 &1 & 0 & Brane \\
$\phi_\mu$ 		& 3'& 1 & 4 & 1 &0 & 0 & Brane \\
$\phi_e$ 			& 3'& 1 & 0 & 1 &1 & 0 & $+1$\\
\rule{0pt}{3ex}%
$A_1$ 			& 1& 1 &  4& 3 &0 & 0 & +1\\
$A_{3'}$ 			& 3'& 1 &  1 &  3 & 0&  1 & Brane \\
$A_{2}$ 			& 2& 1 &  0& 3 &0&  0 & +1\\
$A_{1}'$ 			& 1'& 1 &  2 & 3 &0&  0 & Brane\\
\hline
\end{tabular}
\caption{Superfields in the model with discrete symmetries. }
\label{tab:funfield55}
\end{table}

With a discrete symmetry, extra Yukawa terms are allowed. However any extra term from eq. \ref{eq:yuk} will be highly suppressed and $<O(10^{-10})$. Even though these terms are completely negligible for the flavour observables, they can upset the mechanism that avoids the strong CP problem. These terms would generate an angle $\bar{\theta}<10^{-10}$, that barely satisfies experimental constraints coming from the neutron electric dipole moment \cite{Baker:2006ts}.

In this discrete setup, there can be extra term that allow proton decay. The largest contribution is 
\begin{equation}
T^\pm_1 T^\pm_1 T^\pm_1 F \frac{\braket{\xi^5\phi_e}}{\Lambda^6},
\end{equation}
that complies with the experimental constraint of the terms
\begin{equation}
g\braket{X}TTTF,\hspace{7mm} {\rm with}\hspace{7mm} g\braket{X}<10^9.
\end{equation}
In our case
\begin{equation}
g\braket{X}\sim \gamma^9\sim 10^{-9},
\end{equation}
the constraint can be easily satisfied.

Another difference is that the now the term $\mu\sim \tilde{\xi}^8 M_{GUT}$,  is larger than one obtained with the $U(1)$ but still close to the correct value.

\end{document}